\documentclass[review,3p]{elsarticle}

\usepackage[english]{babel}
\usepackage{amsmath}
\usepackage{amssymb}
\usepackage{dsfont}
\usepackage{graphicx}
\usepackage[colorlinks=true, allcolors=blue]{hyperref}
\usepackage{xurl}
\usepackage{enumitem}
\usepackage{makecell}
\usepackage{bbold}
\usepackage{commath}
\usepackage{float}
\usepackage{longtable}

\usepackage{standalone}
\usepackage{tikz} 
\usepackage{xfp}
\usetikzlibrary{shapes,arrows}

\usepackage[labelsep=period]{caption}
\usepackage{multirow}
\usepackage{multicol}

\usepackage{lineno}

\usepackage{bm}
\usepackage{comment}
\usepackage{pbox}
\usepackage{xcolor}
\usepackage{subcaption}
\graphicspath{ {image/} }

\usepackage{nameref}

\usepackage{titlesec}
\titleformat{\paragraph}[runin]{\normalfont\normalsize\bfseries}{\theparagraph}{1em}{}
\titlespacing*{\paragraph}{0pt}{0pt}{1em}{}

\usepackage{lipsum}
\makeatletter \def\ps@pprintTitle{\let\@oddhead\@empty \let\@evenhead\@empty \def\@oddfoot{\hfill\today} \let\@evenfoot\@oddfoot} \makeatother

\begin{document}

\begin{frontmatter}

\title{A New System Function for Maximum Processable Flow in\\ Process Plants and Application to Reliability Assessment}

\author[1]{Ji-Eun Byun}
\ead{Ji-Eun.Byun@glasgow.ac.uk}
\address[1]{James Watt School of Engineering, University of Glasgow, Glasgow, United Kingdom}

\author[2]{Se-Hyeok Lee\corref{cor1}}
\ead{shlee08@kict.re.kr}
\address[2]{The Korea Institute of Civil Engineering and Building Technology, Republic of Korea}

\cortext[cor1]{Corresponding author}

\begin{abstract}
This study presents a new system performance function for process plant reliability analysis, formulated to capture both structural topology and process sequencing constraints. Built on a modified maximum-flow framework and solved via linear programming, the proposed function efficiently quantifies the maximum feasible flow through a series of interconnected stages. It addresses limitations of existing models—such as fault trees and event trees—that often overlook flow continuity and topological dependencies. The system function is integrated into a reliability assessment framework, enabling the evaluation of system failure probability and reliability-based component importance measures. Application to two benchmark examples, including a gas supply plant with 57 nodes and 102 edges, demonstrates the effectiveness of the proposed system function and the validity of the resulting reliability assessment for risk-informed layout planning. A reconfiguration guided by component importance measures yields up to a 20\% reduction in system failure probability, underscoring the importance of effective equipment and pipeline layout. The proposed framework offers a promising direction for reliability-based management of industrial process facilities.
\end{abstract}


\end{frontmatter}

\section{Introduction}

The reliable operation of process plants is a cornerstone of industrial safety and performance, particularly given their role in producing critical outputs such as chemicals and energy. These systems are subject to a broad spectrum of internal and external hazards, ranging from equipment degradation to extreme events such as earthquakes and fire–explosion hazards. Consequently, decision-makers require analytical tools that rigorously quantify system reliability while accounting for such uncertainties. Reliability-based methodologies have thus emerged as indispensable components in the design and operational planning of process facilities.

A substantial body of work has focussed on risk analysis and mitigation strategies in process industries, particularly under conditions involving cascading or compound hazards. For example, considerable research has been directed toward Natech (Natural Hazard Triggering Technological Disasters) risk assessment, aiming to prevent secondary incidents such as toxic material dispersion and fire-induced domino effects \cite{MesCasMun20, NasHaz16}. Complementing these efforts are industrial guidelines that standardise quantitative risk assessment procedures \cite{Fre90}. From a structural and organisational perspective, facility layout optimisation has been studied as a means of improving resilience through robust connectivity between functional subsystems \cite{PerMulDia21}. Additionally, advanced analytical frameworks—such as state-space models—have been employed to assess safety-critical systems and enumerate failure event sequences with associated probabilities \cite{KumSinTri17}.

Reliability analysis is also integral to decision support on effective system design. It facilitates identification of critical assets, optimisation of system toplogy, and prioritisation of maintenance. For instance, equipment may be ranked by their influence on system-level failure probabilities to guide risk mitigation strategies \cite{MocPil23}, and system architecture may be evaluated for vulnerability using reachability and structural analysis \cite{GouPetCoc20}. Maintenance optimisation has likewise been informed by reliability assessments, as systematically reviewed in \cite{Gac19}, using techniques such as importance measures \cite{VisReg16}, dynamic Bayesian network \cite{OzgTurKar19}, and hierarchical system representations \cite{FerZio14}, to derive cost-effective and risk-informed maintenance schedules. Moreover, reliability-centred system design principles have been developed for enhancing architectural resilience at the planning stage \cite{WatMorWei23}.

Recent studies have focussed on modelling the failure probabilities of individual assets, improving predictive capabilities for critical infrastructure elements such as storage tanks—by accounting for fire-induced domino effects and structural buckling under compound loads \cite{AmiScaCoz24, CheCheMo24}—as well as for piping networks and aging components in nuclear and chemical facilities \cite{BeaReiSak25, MarSanCar23}. These component-level models can be incorporated into system-level reliability frameworks by explicitly modelling the operational dependencies among components. To this end, several modelling paradigms have been proposed. For example, system reliability has been evaluated using series and parallel structures \cite{Kim23}, series-parallel systems \cite{Hsi24}, and $k$-out-of-$n$ models \cite{DraBei24,SuWanDua21}. Additionally, various models have been proposed to consider the non-symmetric roles of system components have also been developed including Markov graphical models for cascading failures in tank farms \cite{LiCheAmy23, BasBeh20}, fault tress and event trees \cite{VelCep24, MenCheYan22,AneUlm18}, Petri nets \cite{YanDunAnd23}, reliability block diagrams \cite{PatDePCos21, LiaYanZha22}, capacity block diagrams \cite{CapKalPao20, CapDonSal23, KalCapCor24}, multilevel flow modeling \cite{SonZhaLin23}, and Bayesian network \cite{KhaKhaAmy13, SegBenMod21, LeeSeoSon24}.

However, these existing models often neglect the operational and topological constraints imposed by process flows. In practice, the throughput of a process plant is dictated not only by the availability of individual assets, but also by the continuity of flows from one processing stage to another—each mediated by a network of pipelines and governed by the system's topology. Ignoring such flow-based and topological constraints can result in misleading assessments of system reliability.

To address this limitation, the present study introduces a new system performance function that evaluates the maximum feasible process flow through a sequence of interconnected stages. The formulation, based on a modified maximum-flow analysis, accounts for both structural topology and process sequencing constraints, and is solved via linear programming to ensure computational efficiency. This system function is then integrated into a reliability assessment framework to quantify the probability of system failure given the failure probabilities of individual components. Leveraging its ability to capture inter-stage and topological dependencies, we further show how reliability-based component importance measures can support effective layout planning. Numerical studies demonstrate that the proposed approach provides a more accurate characterisation of system-level performance than existing reliability assessment methods. They further highlight that system topology significantly influences system reliability even with the identical number of pipelines and equipment.

This paper is organised as follows. Section~\ref{sec:sys_fun} introduces a new formulation to calculate the maximum processable flow, or maximum throughput, of a process plant. The formulation is illustrated using a didactic example and compared with commonly used fault tree–based reliability methods. Section~\ref{sec:rel} demonstrates how the proposed system function can be applied to system reliability assessment and to the computation of reliability-based component importance measures, exemplified in this article using Birnbaum's measures. Section~\ref{sec:ex} presents two numerical examples. Section~\ref{sec:con} concludes the paper, and the datasets used in the numerical examples are provided in \ref{sec:ex_data}.

\section{System Performance Function Based on Linear Programming}\label{sec:sys_fun}

\subsection{System representation and modelling assumptions}

We define the {\it maximum processable flow} of a process plant system as the highest feasible amount of material (e.g. gases and fluids) that can be transmitted from the first-stage nodes to the end-stage nodes, while respecting the sequential nature of the underlying process stages. This flow must traverse all intermediate stages in order, and is constrained by the capacities and operational states of individual assets--specifically, the nodes (e.g. pipeline joints and equipment) and edges (e.g. pipelines). Nodes are referred to as \textit{station} nodes if they serve as points where flows must be held for processing or storage at a given stage. 

$\mathcal{N} = \{1, \ldots, K\}$ denotes the set of node indices. Each node $n_k$ is associated with a maximum capacity $c_k$ that represents the storage capacity for station nodes and the processable throughput for non-station nodes. For each stage $m \in \{ 1, \ldots, M \}$, we denote the set of station nodes for stage~$m$ as $\mathcal{S}_m = \{k : k \in \mathcal{N}_m\}$,
where $\mathcal{N}_m \subset \mathcal{N}$ is the set of station nodes of stage~$m$.
Meanwhile, we denote the set of edges as $\mathcal{E} = \{\left( i, j, m\right)\}$, with $i$ and $j$ referring to the tail and head nodes connected by the edge and $m$ indicating that the edge carries flow from stage~$m$ to stage~$m+1$. A given node pair $(i, j)$ may be associated with multiple stage transitions; that is, the same pair may appear multiple times in $\mathcal{E}$, each associated with a different value of $m$. Additionally, each edge $(i,j,m)$ has capacity $c_{ij}^m$.

\begin{figure}[H]
    \centering

    \begin{subfigure}[t]{0.49\textwidth}
        \centering
        \includegraphics[width=\linewidth]{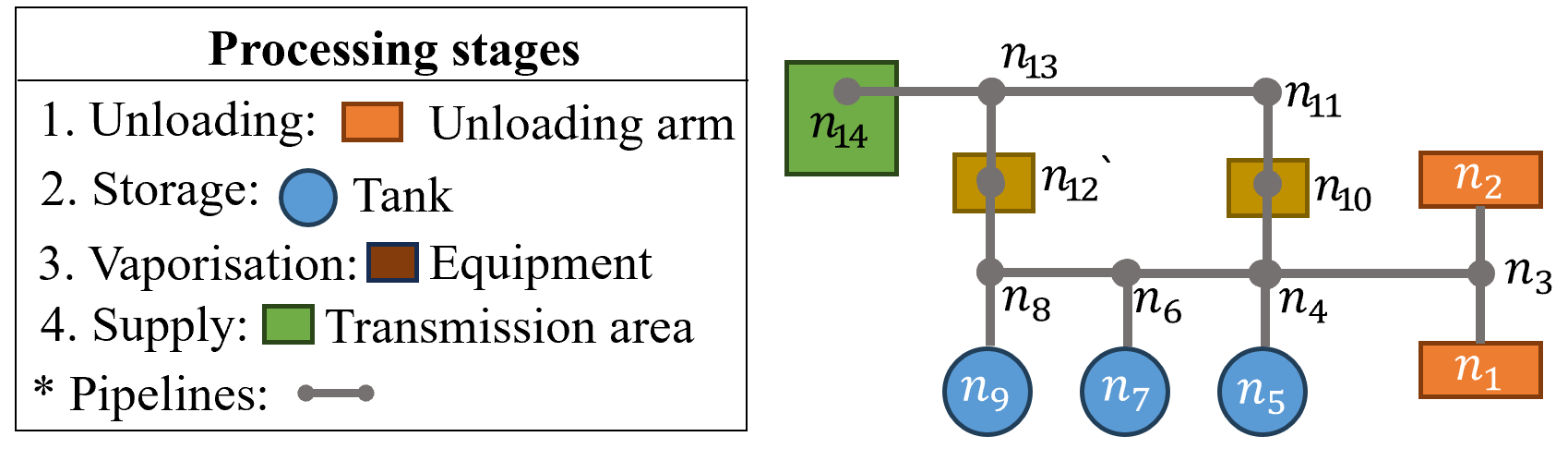}
        \caption{}
        \label{fig:toy_graph}
    \end{subfigure}
    \par\bigskip
    
    \begin{subfigure}[t]{0.3\textwidth}
        \centering
        \includegraphics[width=\textwidth]{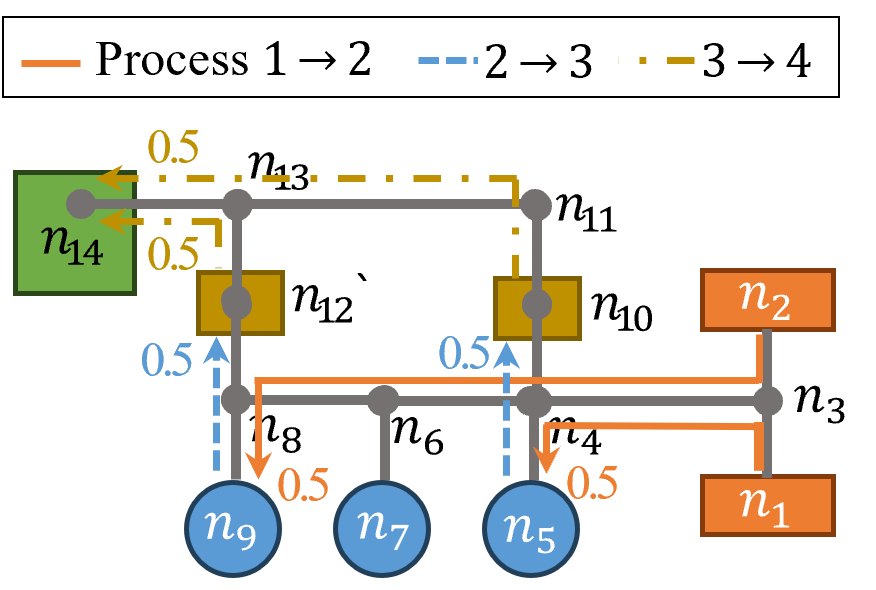}
        \caption{}
        \label{fig:toy_graph_optim1}
    \end{subfigure}
    \hfill
    \begin{subfigure}[t]{0.3\textwidth}
        \centering
        \includegraphics[width=\textwidth]{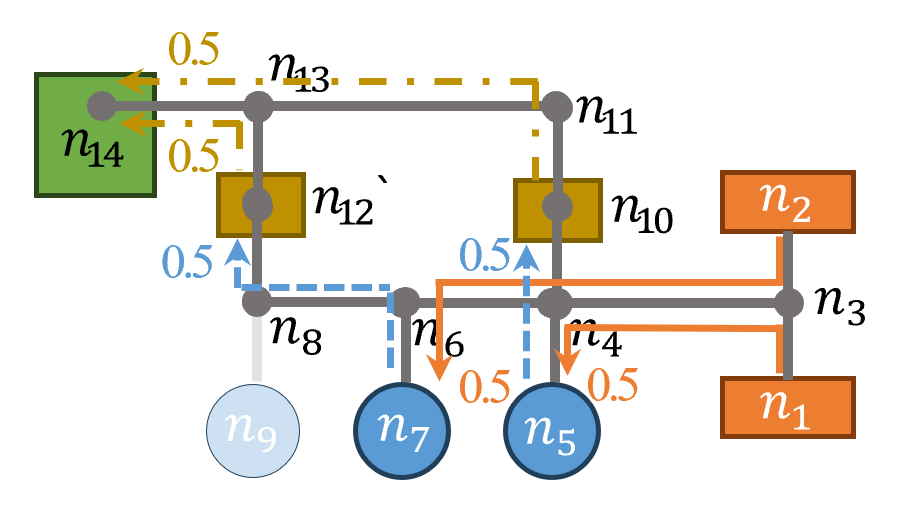}
        \caption{}
        \label{fig:toy_graph_optim2}
    \end{subfigure}
    \hfill
    \begin{subfigure}[t]{0.3\textwidth}
        \centering
        \includegraphics[width=\textwidth]{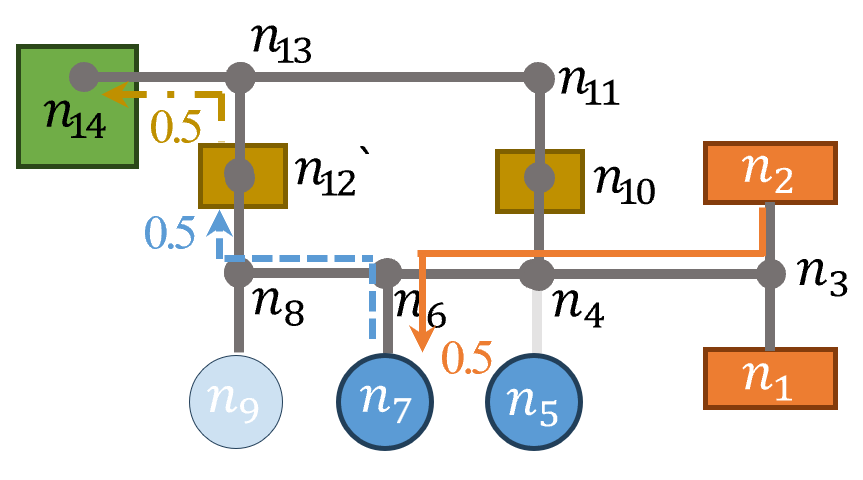}
        \caption{}
        \label{fig:toy_graph_optim3}
    \end{subfigure}

    \caption{Didactic example of a process plant system. (a) Network layout, where flows must sequentially reach process stages 1 through 4, as illustrated in the left box. (b)-(d) show maximum processable flow configurations (note that these configurations are not unique). Flow directions are indicated by line styles: orange solid lines for stages $1\rightarrow 2$, dashed blue lines for stages $2\rightarrow 3$, and dash-dotted brown lines for stages $3\rightarrow 4$. (b) All assets are functional; the maximum processable flow is 1.0. (c) With $n_9$ and $e_9$ failed, the maximum flow remains at 1.0. (d) When $n_9$ and $e_4$ fail, the maximum flow is reduced to 0.5.}
    \label{fig:toy_ex}
\end{figure}

To illustrate this, consider a didactic example shown in Fig.~\ref{fig:toy_graph}. There are four stages that the flow must be held in order, where the station nodes and edges are summarised in Table~\ref{tab:toy_stages} and Table~\ref{tab:toy_edges}, respectively.
Fig.~\ref{fig:toy_graph_optim1}-\ref{fig:toy_graph_optim3} illustrate how the maximum processable flow responds to different failure scenarios of pipelines and equipment. Fig.~\ref{fig:toy_graph_optim1} shows when all assets are functional, in which case the maximum flow is 1.0 with flow configurations shown in the figure. In Fig.~\ref{fig:toy_graph_optim2}, node $n_9$ and edges $(8,9,1)$ and $(9,8,2)$ are assumed to have failed; however, the system retains the maximum flow 1.0 owing to an alternative route via $n_7$ and $(7,6,1)$ and $(7,6,2)$. In contrast, Fig.~\ref{fig:toy_graph_optim3} illustrates the case where node $n_9$ and edges $(4,5,1)$ and $(5,4,2)$ fail. In this case, the only accessible station node in stage 2 is $n_7$, reducing the maximum processable flow to 0.5. These outcomes can be reproduced by the proposed optimisation formulation in the next subsection.

\begin{table}[H]
    \centering
    \begin{tabular}{c|c|c}
        \hline
        Stage & Station node indices & Capacity \\ \hline
        1 & $n_1$, $n_2$ & 0.5 \\
        2 & $n_5$, $n_7$, $n_9$ & 0.5 \\
        3 & $n_{10}$, $n_{12}$ & 0.5 \\
        4 & $n_{14}$ & 1.0 \\ \hline
    \end{tabular}
    \caption{Station nodes of the didactic example. Station nodes of the same stage have the same capacity.}
    \label{tab:toy_stages}
\end{table}

\begin{table}[H]
    \centering
    \begin{tabular}{c|c}
        \hline
        Node pair & Transition start stage \\ \hline
        $(1, 3)$ & 1 \\
        $(2, 3)$ & 1 \\
        $(3, 4)$ & 1 \\
        $(4, 5)$ & 1 \\
        $(5, 4)$ & 2 \\
        $(4, 6)$ & 1, 2 \\
        $(6, 4)$ & 2 \\
        $(4, 10)$ & 2 \\
        $(6, 7)$ & 1 \\
        $(7, 6)$ & 2 \\
        $(6, 8)$ & 1, 2 \\
        $(8, 6)$ & 2 \\
        $(8, 9)$ & 1 \\
        $(9, 8)$ & 2 \\
        $(8, 12)$ & 2 \\
        $(10, 11)$ & 2 \\
        $(11, 12)$ & 3 \\
        $(12, 13)$ & 3 \\
        $(13, 14)$ & 3 \\ \hline
    \end{tabular}
    \caption{Edges of the didactic example. All edges have a capacity of 1.0.}
    \label{tab:toy_edges}
\end{table}

\subsection{Formulation of system performance function}\label{sec:optim}

The maximum processable flow of the system is computed by solving the following linear programming problem, which extends the classical maximum flow formulation to account for the multi-stage structure of process plant systems. The optimisation problem aims to determine the maximum processable flow across a process plant system subject to flow and capacity constraints. The decision variables consist of $\{f_{ij}^m : (i,j,m)\in\mathcal{E}\}$, representing the flow from $n_i$ to $n_j$ during stage $m$, and an auxiliary variable $u$ representing the total flow originated from the first-stage stations.

\begin{subequations}
\begin{align}
    \operatorname*{maximise}_{\left\{f_{ij}^m : \left((i,j), m\right) \in \mathcal{E}\right\} \cup \left\{u\right\}} \quad & u \label{obj} \\ 
    \text{subject to} \quad & \sum_{i\in \mathcal{N}_1, \, j:\left((i,j),1\right)\in\mathcal{E}} f_{ij}^1 = u \label{orig_depot} \\
    & \sum_{j\in \mathcal{N}_m, \, i: \left((i,j), m-1\right)\in\mathcal{E}} f_{ij}^{m} = u, \quad m=2,\ldots,M \label{depots_income} \\
    & \sum_{j : ((j,i), m-1) \in \mathcal{E}} f_{ji}^{m-1} - \sum_{j : ((i,j), m) \in \mathcal{E}} f_{ij}^m = 0, \quad \forall m=2,\ldots,M-1, \quad \forall i \in \mathcal{N}_m \label{depots_balance} \\
    & \sum_{j : ((j,i), m) \in \mathcal{E}} f_{ji}^{m} - \sum_{j : ((i,j), m) \in \mathcal{E}} f_{ij}^{m} = 0, \quad \forall m=1,\ldots,M, \quad \forall i \in \mathcal{N} \setminus \bigcup_{m'=1}^{M} \mathcal{N}_{m'} \label{non_depots_balance} \\
    & 0 \leq f_{ij}^m \leq \max \{c_{ij}^m, c_i, c_j\}, \quad \forall ((i,j),m) \in \mathcal{E} \label{capa}
\end{align}\label{eq:optim}
\end{subequations}

The objective function~\eqref{obj} maximises $u$, corresponding to the total amount of flow initiated at stage 1 by constraint~\eqref{orig_depot}.
Constraint~\eqref{depots_income} enforces that, at each station node associated with stage $m$, $m=2,\ldots,M$, the total inflow matches the total flow $u$, thereby ensuring that the flow is fully transferred through each required stage without loss at station nodes.
Constraint~\eqref{depots_balance} guarantees flow continuity across intermediate station nodes for stages $2$ to $M-1$, such that the total outflow into a station node at stage $m-1$ equals the total inflow from the node at stage $m$.
Constraint~\eqref{non_depots_balance} imposes flow conservation at non-station nodes for all stages; the total inflow and total outflow at any such node must be the same.
Finally, constraint~\eqref{capa} enforces that the flow on each edge remains non-negative and does not exceed its capacity as well as the capacities of its two end nodes.


\subsection{Application to the didactic example}

\subsubsection{Computation of the system performance function}

To illustrate the application of the proposed system function, Table~\ref{tab:toy_edges_topim} summarises the capacities and optimised flows on each edge under the configurations of the didactic example shown in Fig.~\ref{fig:toy_graph_optim1}-\ref{fig:toy_graph_optim3}.

\begin{longtable}{c|cc|cc|cc}
    \hline
    \multirow{2}{*}{$\left(i, j, m\right)$} & \multicolumn{2}{c|}{Fig.~\ref{fig:toy_graph_optim1}} & \multicolumn{2}{c|}{Fig.~\ref{fig:toy_graph_optim2}} & \multicolumn{2}{c}{Fig.~\ref{fig:toy_graph_optim3}} \\ \cline{2-7}
    & $c_{ij}^m$ & $f_{ij}^m$ & $c_{ij}^m$ & $f_{ij}^m$ & $c_{ij}^m$ & $f_{ij}^m$ \\ \hline
    $\left(1, 3, 1\right)$ & 1.0 & 0.5 & 1.0 & 0.5 & 1.0 & 0.0 \\
    $\left(2, 3, 1\right)$ & 1.0 & 0.5 & 1.0 & 0.5 & 1.0 & 0.0 \\
    $\left(3, 4, 1\right)$ & 1.0 & 1.0 & 1.0 & 1.0 & 1.0 & 0.5 \\
    $\left(4, 5, 1\right)$ & 1.0 & 0.5 & 1.0 & 0.5 & 0.0 & 0.0 \\
    $\left(5, 4, 2\right)$ & 1.0 & 0.5 & 1.0 & 0.5 & 0.0 & 0.0 \\
    $\left(4, 6, 1\right)$ & 1.0 & 0.5 & 1.0 & 0.5 & 1.0 & 0.5 \\
    $\left(4, 6, 2\right)$ & 1.0 & 0.0 & 1.0 & 0.0 & 1.0 & 0.0 \\
    $\left(6, 4, 2\right)$ & 1.0 & 0.0 & 1.0 & 0.0 & 1.0 & 0.0 \\
    $\left(4, 10, 2\right)$ & 1.0 & 0.5 & 1.0 & 0.5 & 1.0 & 0.0 \\
    $\left(6, 7, 1\right)$ & 1.0 & 0.0 & 1.0 & 0.5 & 1.0 & 0.5 \\
    $\left(7, 6, 2\right)$ & 1.0 & 0.0 & 1.0 & 0.5 & 1.0 & 0.5 \\
    $\left(6, 8, 1\right)$ & 1.0 & 0.5 & 1.0 & 0.0 & 1.0 & 0.0 \\
    $\left(6, 8, 2\right)$ & 1.0 & 0.0 & 1.0 & 0.5 & 1.0 & 0.5 \\
    $\left(8, 6, 2\right)$ & 1.0 & 0.0 & 1.0 & 0.0 & 1.0 & 0.0 \\
    $\left(8, 9, 1\right)$ & 1.0 & 0.5 & 0.0 & 0.0  & 1.0 & 0.0 \\
    $\left(9, 8, 2\right)$ & 1.0 & 0.5 & 0.0 & 0.0  & 1.0 & 0.0 \\
    $\left(8, 12, 2\right)$ & 1.0 & 0.5 & 1.0 & 0.5 & 1.0 & 0.5 \\
    $\left(10, 11, 2\right)$ & 1.0 & 0.5 & 1.0 & 0.5 & 1.0 & 0.0 \\
    $\left(11, 12, 3\right)$ & 1.0 & 0.5 & 1.0 & 0.5 & 1.0 & 0.0 \\ 
    $\left(12,13, 3\right)$ & 1.0 & 0.5 & 1.0 & 0.5 & 1.0 & 0.5 \\
    $\left(13,14, 3\right)$ & 1.0 & 0.5 & 1.0 & 0.5 & 1.0 & 0.5 \\ \hline
\caption{Optimisation results for the didactic example under the three scenarios illustrated in Fig.~\ref{fig:toy_graph_optim1}, Fig.~\ref{fig:toy_graph_optim2}, and Fig.~\ref{fig:toy_graph_optim3}, respectively. Each row corresponds to an edge $\left(i,j, m\right)$, where $c_{ij}^m$ denotes the effective capacity used in constraint~\eqref{capa}, and $f_{ij}^m$ represents the optimised decision variable obtained by solving the system performance function in \eqref{eq:optim}.}
\label{tab:toy_edges_topim}
\end{longtable}

\subsubsection{Comparison with fault tree-based reliability analysis}

\begin{figure}[H]
    \centering
    \includegraphics[width=0.6\linewidth]{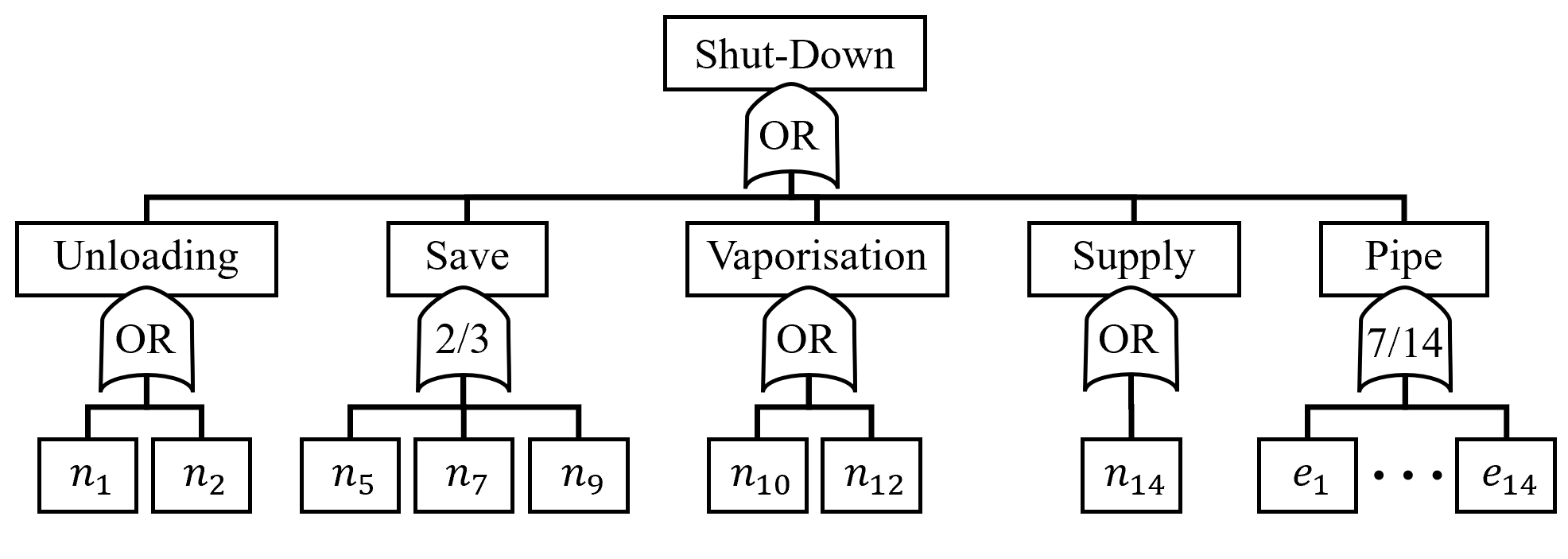}
    \caption{Fault tree representation of the didactic example. The system is considered to have failed if the maximum processable flow capacity falls below 1.0. The treatment of all assets within the same stage as symmetric in the fault tree model highlights its limitation in capturing their topological differences.}
    \label{fig:toy_fault_tree}
\end{figure}

Fault tree models are commonly used for risk assessment in process plant systems. Revisiting the didactic example, we define system failure as occurring when the maximum processable flow capacity falls below 1.0 (note that the maximum capacity under fully functional conditions is 1.0). Based on this criterion, a corresponding fault tree can be constructed, as shown in Fig.~\ref{fig:toy_fault_tree} (for further details on fault tree modelling in process plant systems, see \cite{LeeSeoSon24}).

In the fault tree, the system fails if any of the four processing stages—unloading, storage, vaporisation, or supply—or the pipe subsystem fails. For the unloading and vaporisation stages, each contains two nodes with capacities of 0.5. Since the failure of any single node prevents the stage from fulfilling the target flow of 1.0, these stages are modelled using OR gates. The storage stage comprises three nodes, each also with a capacity of 0.5, and is represented as a 2-out-of-3:F system. Here, a $k$-out-of-$N$:F system denotes a system that fails if $k$ or more out of $N$ components fail. The supply stage contains only one node and thus fails if that single node fails. For the pipe subsystem, the relationship between individual pipeline failures and the system’s overall capacity is less direct—an issue common in many system analyses (cf. \cite{LeeSeoSon24}). In this example, we approximate the subsystem using a 7-out-of-14:F model, assuming the system fails if more than half of the pipelines fail.

This fault tree model highlights a key limitation when compared to the proposed system performance function: it does not account for the topological configuration of pipelines and equipment. For instance, the fault tree cannot differentiate between the two scenarios illustrated in Figs.~\ref{fig:toy_graph_optim2} and \ref{fig:toy_graph_optim3}. In both cases, one node in the storage stage and one pipeline are failed, so the fault tree concludes the system survives in both scenarios. However, the proposed system function yields different outcomes: in Fig.~\ref{fig:toy_graph_optim2}, the processable flow remains at 1.0, whereas in Fig.~\ref{fig:toy_graph_optim3}, it is reduced to 0.5—leading to system failure. This distinction is possible because the proposed optimisation captures the differing topological impact of individual assets.

Although fault trees are often converted into Bayesian network to enable more detailed probabilistic inference, this transformation does not address the underlying limitation: the loss of spatial and structural information remains inherent to the abstraction.

\section{Application to system reliability analysis}\label{sec:rel}

\subsection{Incorporating uncertainty in asset capacities}

To reflect the random operational states of assets, the capacity variables of nodes, $c_k$, $k\in\mathcal{N}$, and edges, $c_{ij}^m$, $\left( i,j,m \right) \in \mathcal{E}$ can be adjusted. For instance, when each asset is either functional or dysfunctional, its state can be represented by a binary random variable $X_n$, $n = 1, \ldots, N$, where $X_n = 1$ indicates survival and $X_n = 0$ indicates failure. For reliability assessment, a joint distribution $P(\bm{X})$ of a set of component random variables $\bm{X} = \{X_1, \ldots, X_N\}$ needs to be defined; if the random variables are independent, this reduces to $P(\bm{X}) = \prod_n P(X_n)$. As a component random variable represents a physical asset, some $X_n$ may correspond to one or more nodes in $\mathcal{N}$ and/or edges in $\mathcal{E}$. For instance, in the didactic example, edges $(4,6,1)$, $(4,6,2)$, and $(6,4,2)$ may be assigned the same random variable if the transitions between $n_4$ and $n_6$ are realised by the same physical asset.

As an illustration, consider the binary-state case. Given an assignment of $\bm{X}$ (i.e. a realisation of states of $X_1,\ldots,X_N$), the capacity $c_k$ or $c_{ij}^m$ of an asset associated with $X_n$ is set to zero if $X_n = 0$, and to its original value if $X_n = 1$. The resulting capacities can then be used to specify Constraint~\eqref{capa} in the proposed system function.

\subsection{Quantifying component importance for effective layout planning}\label{sec:cim}

The proposed system function emphasises the importance of the \textit{topological} roles of individual assets. In other words, the system function may return different outcomes even with the same number of pipelines and equipment, depending on how they are arranged. Reliability assessment using this system function can therefore further support effective layout planning through the evaluation of reliability-based component importance measures, such as Birnbaum's measure \cite{ScherbGarreStraub19} and the conditional probability-based importance measure \cite{KangSongGardoni08}. These measures provide insights by quantifying how changes in a component random variable's survival or failure probability affect overall system reliability. By comparing importance values, the associated assets can be prioritised to inform expansion.

For example, Birnbaum's measure of a component random variable $X_n$ is defined as
\begin{linenomath*}
\begin{equation}
    BI_n = P(S=1 \mid X_n=1) - P(S=1 \mid X_n=0), \ n=1,\ldots,N,
\end{equation}\label{eq:bm}
\end{linenomath*}
where $S$ is a random variable whose state 0 represents system failure and 1 system survival.
The measure can be evaluated via two system reliability analyses: one assuming a modified probability distribution with $P(X_n=1)=1$ and the other with $P(X_n=1)=0$. 
While system reliability can be evaluated by various approaches, the following numerical examples estimate it using $10^7$ Monte Carlo (MC) samples.

\section{Numerical examples}\label{sec:ex}

\subsection{Pressure regularisation system}\label{sec:ex_pressure}

\begin{figure}[h!]
    \centering
    \includegraphics[width=\linewidth]{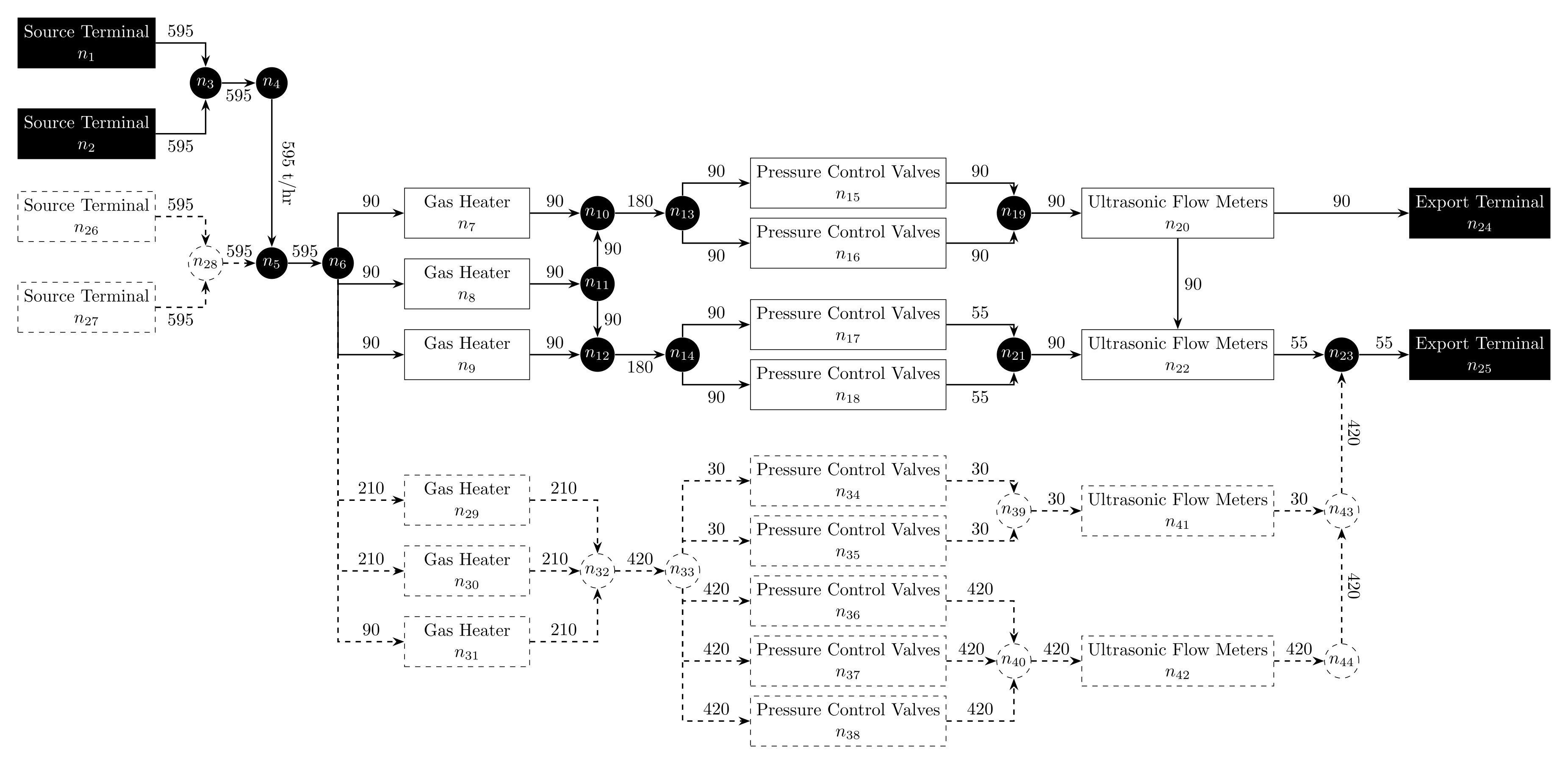}
    \caption{Benchmark pressure regularisation system. The dashed nodes and edges are planned expansion of the plant system. The expansion reduces the system failure probability from 0.233 to 0.183.}
    \label{fig:pressure_layout}
\end{figure}

This section investigates a benchmark pressure regularisation system, which is based on an actual plant layout, with details anonymised for confidentiality. The layout is illustrated in Fig.~\ref{fig:pressure_layout}, where the dashed nodes and edges represent a planned expansion. The system originally comprises 25 nodes and 29 edges, while the planned expansion increases this to 44 nodes and 55 edges. The process involves five stages in order: source terminal (stage 1), gas heater (stage 2), pressure control valves (stage 3), ultrasonic flow meters (stage 4), and export terminal (stage 5). With edges representing pipelines for transport, random variables are assigned to each of station nodes and edges. This leads to 42 random variables for the original layout and 80 for the expanded one. For demonstration, all random variables are assigned a failure probability of 0.03, i.e. $P(X_n=0)$ for $n=1,\ldots,80$, and are assumed statistically independent, i.e. $P(\bm{X})=\Pi_n P(X_n)$.\footnote{For more detailed fragility analyses of plant assets, readers are referred to \cite{LeeSeoSon24} which illustrates quantification of failure probabilities of different asset types under the scenarios of earthquake and earthquake-induced fires and explosions.} The summary of nodes, edges, and random variables is available in Section~\ref{sec:pressure_data}.

The original and expanded layouts have a maximum throughput of 145 t/hr ($=90+55$), with the export terminal acting as the bottleneck. We set the target system throughput as 90 t/hr. System failure probability is estimated using $10^7$ MC samples, which is computationally feasible given that each evaluation of the proposed linear-programming-based system function requires only fractional seconds. Under the system target, the original layout exhibits a failure probability of 23.3\%, which is reduced to 18.3\% in the expanded layout. Notably, despite having little effect on the maximum throughput, the expanded layout leads to a notable improvement in system reliability, underscoring the critical role of plant system topology in reliability management.

\subsection{Gas supply plant}\label{sec:gas}

\begin{figure}[H]
    \centering

    \begin{subfigure}[t]{0.45\textwidth}
        \centering
        \includegraphics[width=\linewidth]{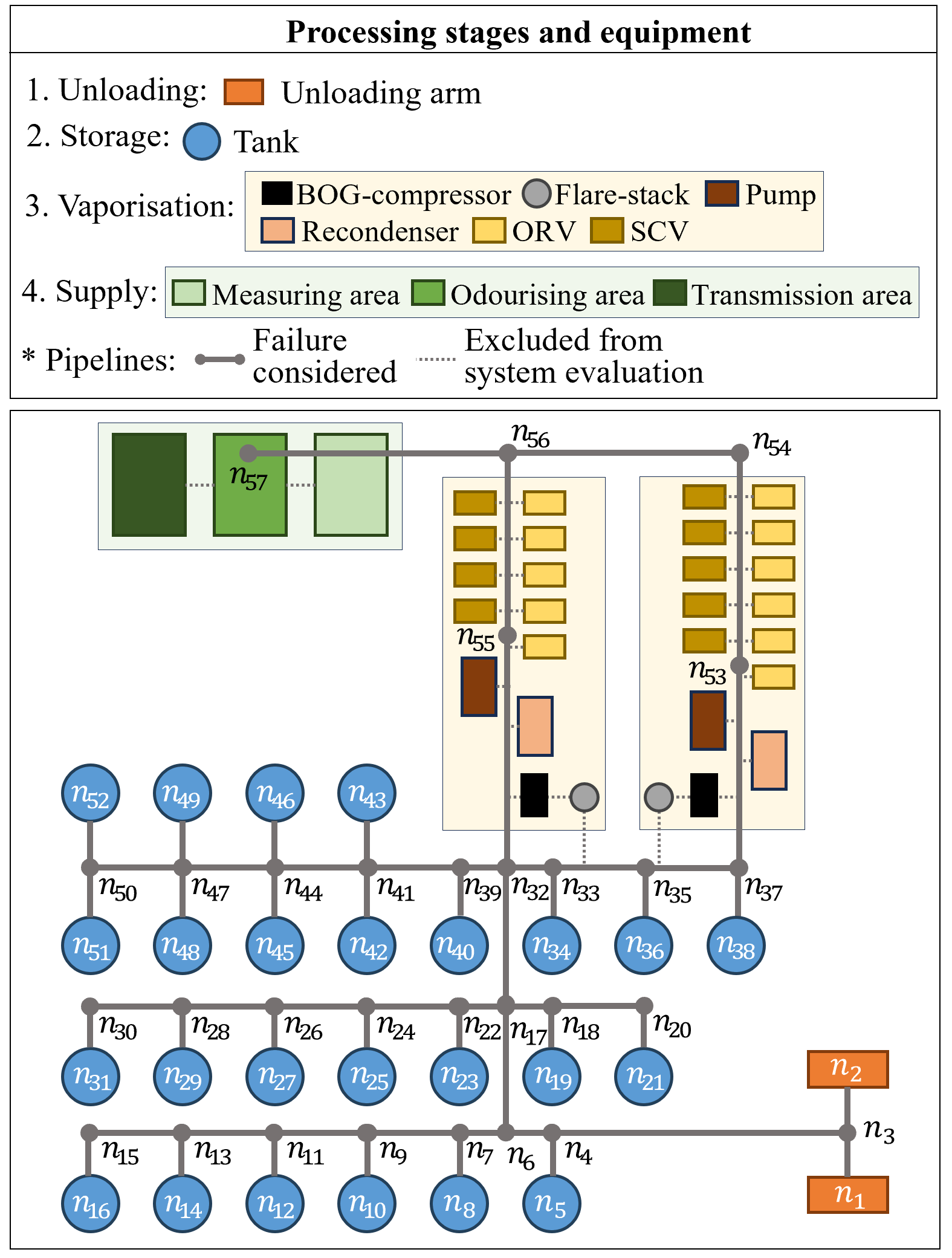}
        \caption{}
        \label{fig:full_ex}
    \end{subfigure}
    \hfill
    \begin{subfigure}[t]{0.45\textwidth}
        \centering
        \includegraphics[width=\linewidth]{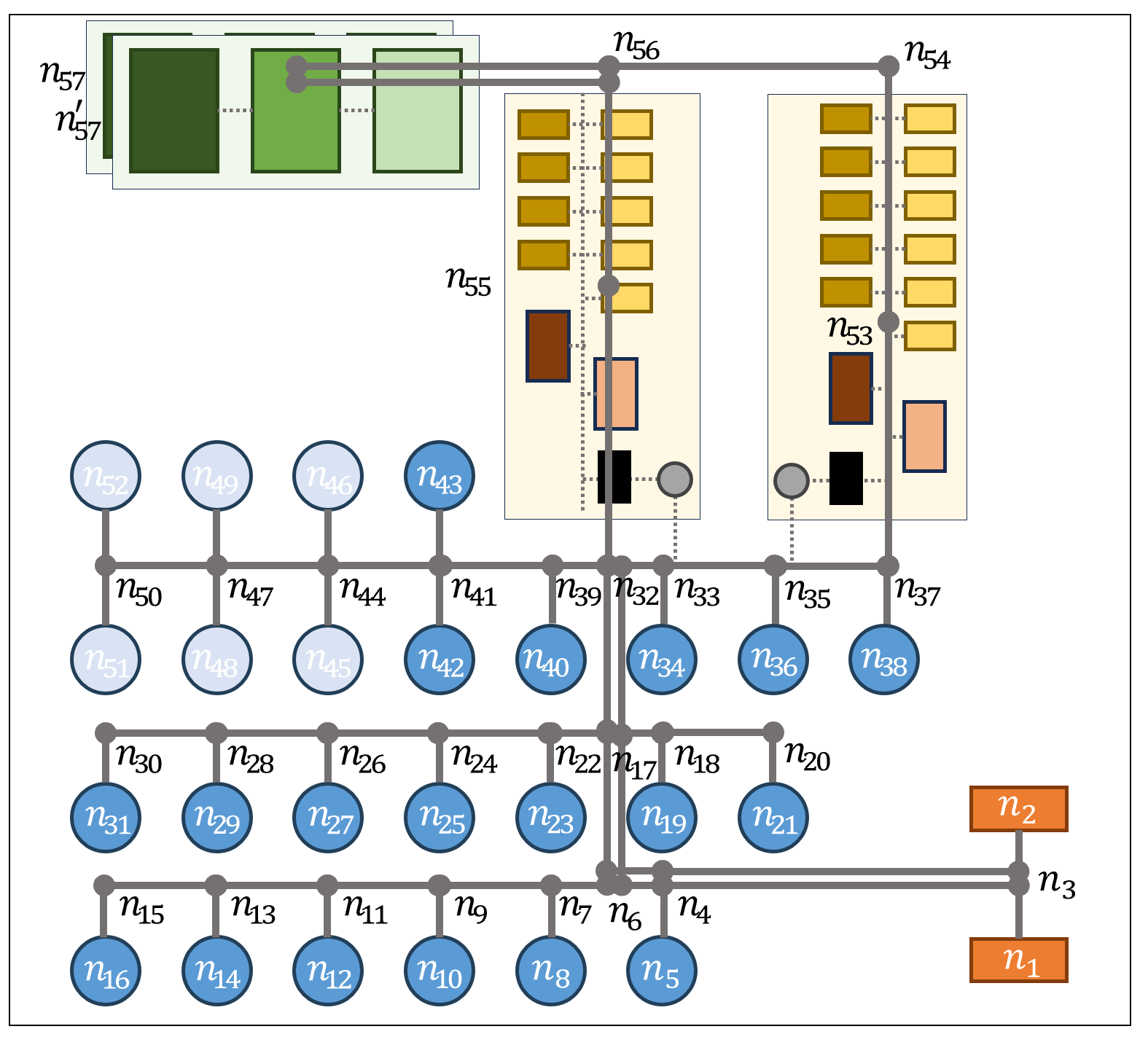}
        \caption{}
        \label{fig:improved_ex}
    \end{subfigure}

    \caption{Benchmark gas supply system adapted from \cite{LeeSeoSon24}. (a) Original topology. (b) Improved topology obtained by adding the six assets with the highest Birnbaum's measures and the six with the lowest values. This new topology reduces the system failure probability from 0.229 to 0.184, reducing it by around 20\%.}
    \label{fig:ex_comparison}
\end{figure}

This section investigates a benchmark gas supply plant adapted from \cite{LeeSeoSon24}, which is shown in Fig.~\ref{fig:full_ex}. The system comprises 57 nodes and 102 edges and involves four sequential processes: unloading (stage 1), storage (stage 2), vaporisation (stage 3), and supply (stage 4). Similarly to the previous example, a random variable is assigned to each of the station nodes and edges representing pipelines, while the same random variable is assigned to edges that connect the same end nodes and only differ in transition stages. This yields 87 random variables. For demonstration, all random variables are assigned a failure probability of 3\%, i.e. $P(X_n=0)=0.03$, $n=1,\ldots,87$, and assumed statistically independent, i.e. $P(\bm{X}) = \Pi_n P(X_n)$. The summary of nodes, edges, and random variables is presented in Section~\ref{sec:gas_data}.

The maximum throughput of the system is 1.0, and the target flow is set to 0.5. Using $10^7$ MC samples, the system failure probability is estimated as 22.9\%. To assess the relative importance of random variables, we compute the Birnbaum's measures in Eq.~\eqref{eq:bm} of $X_1,\ldots,X_{87}$, where the two probabilities in the formulation are estimated by $10^7$ MC samples. The random variables with the highest Birnbaum's measures are listed in Table~\ref{tab:gas_high_bm}. 

As summarised in the table, the random variables with the highest tier, with Birnbaum's measures of approximately 0.84, correspond to either a unique station node--namely $n_{57}$, which is the only station node in the final stage--or corridor-defining edges $(3,4)$, $(4,6)$, $(6,17)$, $(17,32)$, $(47,48)$, whose failure directly results in system failure. 
The second tier, with Birnbaum's measures of around 0.15, comprises assets that are either topologically critical station nodes--namely $n_{55}$, which is located in closer to the supply stage than the altnerative station node $n_{53}$ in the vaporisation stage--or edges incident to this critical node--namely $(32,55)$ and $(55,56)$. The third tier, with Birnbaum's measures between 0.04 and 0.06, represent assets for which alternative paths exist but are limited in number. These comprise the two station nodes in the unloading stage, $n_1$ and $n_2$ and the further station node of the vaporisation stage, $n_{53}$, or the edges connected to these nodes. The lowest Birnbaum's measures are obtained for the highly redundant station nodes in the storage stage and their connecting edges.

Based on this observation, we design a new topology by adding the six top-tier assets and removing the six lowest-ranked ones, as illustrated in Fig.~\ref{fig:improved_ex}. This results in an estimated system failure probability of 18.4\%, which represents a 20\% reduction compared with the original configuration.

\begin{table}[h!]
    \centering
    \begin{tabular}{c|l|l|l}
    \hline
        \makecell[c]{Birnbaum\\[-5pt]measure} & \makecell[c]{Random variables} & \makecell[c]{Associated\\[-5pt]nodes and edges} & \makecell[c]{Topological roles} \\ \hline
        $\sim0.84$ & \makecell[l]{$X_{30}$, $X_{33}$, $X_{35}$,\\[-5pt]$X_{46}$, $X_{61}$, $X_{87}$} & \makecell[l]{$n_{57}$, $(3, 4)$, $(4,6)$,\\[-5pt]$(6,17)$, $(17,32)$, $(56, 57)$} & \makecell[l]{Unique station nodes or\\[-5pt]corridor edges} \\
        $\sim0.15$ & $X_{29}$, $X_{83}$, $X_{86}$ & $n_{55}$, $(32,55)$, $(55,56)$ & \makecell[l]{Topologically critical station nodes or\\[-5pt]their connecting edges} \\
        0.04--0.06 & \makecell[l]{$X_{1}$, $X_{2}$, $X_{28}$,\\[-5pt]$X_{31}$, $X_{32}$, $X_{62}$,\\[-5pt]$X_{64}$, $X_{66}$, $X_{82}$,\\[-5pt]$X_{84}$, $X_{85}$} & \makecell[l]{$n_1$, $n_1$, $n_{53}$,\\[-5pt]$(1,3)$, $(2,3)$, $(32,33)$,\\[-5pt]$(33,35)$, $(35,37)$, $(37,53)$,\\[-5pt]$(53,54)$, $(54,56)$} & \makecell[l]{Station nodes with limited alternatives or\\[-5pt]their connecting edges} \\ \hline
        
    \end{tabular}
    \caption{Random variables with the highest Birnbaum's measures, together with their associated nodes and edges and topologicial roles.}
    \label{tab:gas_high_bm}
\end{table}

\section{Concluding remarks}\label{sec:con}

We propose a new system performance function for process plant systems that quantifies the maximum processable flow through a sequence of processing stages. The function is formulated as a linear programming problem, ensuring computational efficiency even for large-scale systems with hundreds of pipelines and equipment. Unlike conventional fault tree approaches, the proposed function accounts for the topological arrangement of assets and their ability to support flow continuity across stages. This allows it to distinguish the system-level impact of failures of individual assets more accurately.

The proposed system function is readily applicable to reliability analysis by incorporating asset failures through adjustments to the capacity constraints of edges. To further support layout planning, we show how the system function can be used to compute reliability-based component importance measures, with Birnbaum's measure as an illustrative example. By quantifying the relative influence of component random variables on system reliability, these measures support the prioritisation of system expansion.

The proposed method is applied to two benchmark plant systems: a pressure regularisation system, comprising 44 nodes and 55 edges, and a gas supply system, comprising 57 nodes and 102 edges. System reliability analysis is performed using $10^7$ Monte Carlo samples, leveraging the very low computational cost of the proposed formulation, which requires only fractional seconds for both examples. Birnbaum’s measures of the random variables are then computed to identify the most and least critical assets. The results further show that improved system configurations can lead to a notable reduction in system failure probability even with identical numbers of pipelines and equipment.

The proposed system function and its integration with system reliability analysis open new opportunities for reliability-informed management of process plant systems. Future research directions include coupling the framework with reliability-based optimisation to automate topology design. The approach can also be integrated with continuous structural health monitoring, enabling system reliability to be continuously updated and acted upon.

\section*{Acknowledgments}

\noindent This research was supported by the Korea Agency for Infrastructure Technology Advancement (KAIA) grant funded by the Ministry of Land, Infrastructure and Transport (Grant No. RS-2021-KA163162).

\section*{Data availability}
\noindent The datasets used in the numerical examples are available at \url{https://github.com/jieunbyun/network-datasets/tree/main/datasets/process_plants}. The Python code used for the reliability assessment is available at \url{https://github.com/jieunbyun/tsum/tree/main/demos/process_plant}.

\bibliographystyle{elsarticle-num-names} 
\bibliography{refs}

\appendix
\setcounter{table}{0}
\renewcommand{\thetable}{A\arabic{table}}

\section{Numerical example data}\label{sec:ex_data}

\subsection{Pressure regularisation system}\label{sec:pressure_data}

Table~\ref{tab:node_pressure}, Table~\ref{tab:edge_pressure}, and Table~\ref{tab:rv_pressure} respectively summarise the station nodes, edges, and random variables of the pressure regularisation system investigated in Section~\ref{sec:ex_pressure}

\begin{table}[H]
    \centering
    \begin{tabular}{c|c|c|c}
        \hline
        Stage & Status & Station node indices & Capacity (t/hr) \\ \hline
        \multirow{2}{*}{1} & Original & $n_1$, $n_2$ & 595 \\
        & Expansion & $n_{26}$, $n_{27}$ & 595 \\
        \hline
        \multirow{3}{*}{2} & Original & \makecell[c]{$n_7$, $n_8$, $n_9$} & 90 \\
        & Expansion & $n_{29}$, $n_{30}$ & 210 \\
        & Expansion & $n_{31}$ & 90 \\
        \hline
        \multirow{3}{*}{3} & Original & \makecell[c]{$n_{15}$, $n_{16}$, $n_{17}$, $n_{18}$}
          & 90 \\
          & Expansion & $n_{34}$, $n_{35}$ & 30 \\
          & Expansion & $n_{36}$, $n_{37}$, $n_{38}$ & 420 \\
        \hline
        \multirow{3}{*}{4} & Original & $n_{20}$, $n_{22}$ 
          & 110 \\
        & Expansion & $n_{41}$ & 30 \\
        & Expansion & $n_{42}$ & 420 \\
        \hline
        \multirow{2}{*}{5} & Original & $n_{24}$ & 90 \\ 
        & Original & $n_{25}$ & 55 \\
        \hline
    \end{tabular}

    \caption{Station nodes, including planned expansion nodes, of the pressure regularisation benchmark system. For non-station nodes, capacity is defined as the maximum capacity among the connected edges.}
    \label{tab:node_pressure}
\end{table}

\begin{table}[H]
    \centering
    \begin{tabular}{c|c|c||c|c|c}
    \hline
    Edge index & ($i$, $j$, $m$) & Capacity (t/hr) &
    Edge index & ($i$, $j$, $m$) & Capacity (t/hr) \\
    \hline
    $e_1$  & (1, 3, 1)   & 595 & $e_2$  & (2, 3, 1)   & 595 \\
    $e_3$  & (3, 4, 1)   & 595 & $e_4$  & (4, 5, 1)   & 595 \\
    $e_5$  & (5, 6, 1)   & 595 & $e_6$  & (6, 7, 1)   & 90  \\
    $e_7$  & (6, 8, 1)   & 90  & $e_8$  & (6, 9, 1)   & 90  \\
    $e_9$  & (7, 10, 2)  & 90  & $e_{10}$ & (8, 11, 2) & 90  \\
    $e_{11}$ & (9, 12, 2) & 90 & $e_{12}$ & (10, 13, 2) & 180 \\
    $e_{13}$ & (11, 10, 2) & 90 & $e_{14}$ & (11, 12, 2) & 90 \\
    $e_{15}$ & (12, 14, 2) & 180 & $e_{16}$ & (13, 15, 2) & 90 \\
    $e_{17}$ & (13, 16, 2) & 90 & $e_{18}$ & (14, 17, 2) & 90 \\
    $e_{19}$ & (14, 18, 2) & 90 & $e_{20}$ & (15, 19, 3) & 90 \\
    $e_{21}$ & (16, 19, 3) & 90 & $e_{22}$ & (17, 21, 3) & 55 \\
    $e_{23}$ & (18, 21, 3) & 55 & $e_{24}$ & (19, 20, 3) & 90 \\
    $e_{25}$ & (20, 22, 4) & 90 & $e_{26}$ & (20, 24, 4) & 90 \\
    $e_{27}$ & (21, 22, 3) & 90 & $e_{28}$ & (22, 23, 4) & 55 \\
    $e_{29}$ & (23, 25, 4) & 55 & $e_{30}$ & (26, 28, 1) & 595 \\
    $e_{31}$ & (27, 28, 1) & 595 & $e_{32}$ & (28, 5, 1) & 595 \\
    $e_{33}$ & (6, 29, 1) & 210 & $e_{34}$ & (6, 30, 1) & 210 \\
    $e_{35}$ & (6, 31, 1) & 90 & $e_{36}$ & (29, 32, 2) & 210 \\
    $e_{37}$ & (30, 32, 2) & 210 & $e_{38}$ & (31, 32, 2) & 90 \\
    $e_{39}$ & (32, 33, 2) & 420 & $e_{40}$ & (33, 34, 2) & 30 \\
    $e_{41}$ & (33, 35, 2) & 30 & $e_{42}$ & (33, 36, 2) & 420 \\
    $e_{43}$ & (33, 37, 2) & 420 & $e_{44}$ & (33, 38, 2) & 420 \\
    $e_{45}$ & (34, 39, 3) & 30 & $e_{46}$ & (35, 39, 3) & 30 \\
    $e_{47}$ & (36, 40, 3) & 420 & $e_{48}$ & (37, 40, 3) & 420 \\
    $e_{49}$ & (38, 40, 3) & 420 & $e_{50}$ & (39, 41, 3) & 30 \\
    $e_{51}$ & (40, 42, 3) & 420 & $e_{52}$ & (41, 43, 4) & 30 \\
    $e_{53}$ & (42, 44, 4) & 420 & $e_{54}$ & (43, 23, 4) & 420 \\
    $e_{55}$ & (44, 43, 4) & 420 & & & \\ \hline
    \end{tabular}

    \caption{Edges of the pressure regularisation benchmark system. $e_{30}$--$e_{55}$ are edges planned for expansion. $i$, $j$, and $m$ denote start node index, end node index, and start station index, respectively.}
    \label{tab:edge_pressure}
\end{table}

\begin{table}[H]
    \centering
    \begin{tabular}{c|c||c|c||c|c}
    \hline
    \makecell[c]{Random\\[-5pt]variable} & \makecell[c]{Asset\\[-5pt]index} &
    \makecell[c]{Random\\[-5pt]variable} & \makecell[c]{Asset\\[-5pt]index} &
    \makecell[c]{Random\\[-5pt]variable} & \makecell[c]{Asset\\[-5pt]index} \\ \hline
    $X_1$ & $n_1$ & $X_2$ & $n_2$ & $X_3$ & $n_7$ \\
    $X_4$ & $n_8$ & $X_5$ & $n_9$ & $X_6$ & $n_{15}$ \\
    $X_7$ & $n_{16}$ & $X_8$ & $n_{17}$ & $X_9$ & $n_{18}$ \\
    $X_{10}$ & $n_{20}$ & $X_{11}$ & $n_{22}$ & $X_{12}$ & $n_{24}$ \\
    $X_{13}$ & $n_{25}$ & $X_{14}$ & $e_1$ & $X_{15}$ & $e_2$ \\
    $X_{16}$ & $e_3$ & $X_{17}$ & $e_4$ & $X_{18}$ & $e_5$ \\
    $X_{19}$ & $e_6$ & $X_{20}$ & $e_7$ & $X_{21}$ & $e_8$ \\
    $X_{22}$ & $e_9$ & $X_{23}$ & $e_{10}$ & $X_{24}$ & $e_{11}$ \\
    $X_{25}$ & $e_{12}$ & $X_{26}$ & $e_{13}$ & $X_{27}$ & $e_{14}$ \\
    $X_{28}$ & $e_{15}$ & $X_{29}$ & $e_{16}$ & $X_{30}$ & $e_{17}$ \\
    $X_{31}$ & $e_{18}$ & $X_{32}$ & $e_{19}$ & $X_{33}$ & $e_{20}$ \\
    $X_{34}$ & $e_{21}$ & $X_{35}$ & $e_{22}$ & $X_{36}$ & $e_{23}$ \\
    $X_{37}$ & $e_{24}$ & $X_{38}$ & $e_{25}$ & $X_{39}$ & $e_{26}$ \\
    $X_{40}$ & $e_{27}$ & $X_{41}$ & $e_{28}$ & $X_{42}$ & $e_{29}$ \\
    $X_{43}$ & $n_{26}$ & $X_{44}$ & $n_{27}$ & $X_{45}$ & $n_{29}$ \\
    $X_{46}$ & $n_{30}$ & $X_{47}$ & $n_{31}$ & $X_{48}$ & $n_{34}$ \\
    $X_{49}$ & $n_{35}$ & $X_{50}$ & $n_{36}$ & $X_{51}$ & $n_{37}$ \\
    $X_{52}$ & $n_{38}$ & $X_{53}$ & $n_{41}$ & $X_{54}$ & $n_{42}$ \\
    $X_{55}$ & $e_{30}$ & $X_{56}$ & $e_{31}$ & $X_{57}$ & $e_{32}$ \\
    $X_{58}$ & $e_{33}$ & $X_{59}$ & $e_{34}$ & $X_{60}$ & $e_{35}$ \\
    $X_{61}$ & $e_{36}$ & $X_{62}$ & $e_{37}$ & $X_{63}$ & $e_{38}$ \\
    $X_{64}$ & $e_{39}$ & $X_{65}$ & $e_{40}$ & $X_{66}$ & $e_{41}$ \\
    $X_{67}$ & $e_{42}$ & $X_{68}$ & $e_{43}$ & $X_{69}$ & $e_{44}$ \\
    $X_{70}$ & $e_{45}$ & $X_{71}$ & $e_{46}$ & $X_{72}$ & $e_{47}$ \\
    $X_{73}$ & $e_{48}$ & $X_{74}$ & $e_{49}$ & $X_{75}$ & $e_{50}$ \\
    $X_{76}$ & $e_{51}$ & $X_{77}$ & $e_{52}$ & $X_{78}$ & $e_{53}$ \\
    $X_{79}$ & $e_{54}$ & $X_{80}$ & $e_{55}$ & & \\ \hline
    \end{tabular}

    \caption{Random variables of the pressure regularisation benchmark system. $X_{43}$--$X_{80}$ correspond to planned pipelines and equipment for expansion. For demonstration, all random variables have an identical failure probability as 3\%, i.e. $P(X_n=0)=0.03$, $n=1,\ldots,80$.}
    \label{tab:rv_pressure}
\end{table}

\subsection{Gas supply plant}\label{sec:gas_data}

Table~\ref{tab:ex_stages}, Table~\ref{tab:edge_gas}, and Table~\ref{tab:rv_gas} summarise respectively the station nodes, edges, and random variables of the gas supply plant system investigated in Section~\ref{sec:gas}.

\begin{table}[H]
    \centering
    \begin{tabular}{c|c|c}
        \hline
        Stage & Station node indices & Capacity \\ \hline
        1 & $n_1$, $n_2$ & 0.5 \\
        2 & \makecell[c]{$n_5$, $n_8$, $n_{10}$, $n_{12}$, $n_{14}$, $n_{16}$, $n_{19}$, $n_{21}$, $n_{23}$, $n_{25}$,\\[-5pt]$n_{27}$, $n_{29}$, $n_{31}$, $n_{34}$, $n_{36}$, $n_{38}$, $n_{40}$, $n_{42}$, $n_{43}$, $n_{45}$,\\[-5pt]$n_{46}$, $n_{48}$, $n_{49}$, $n_{51}$, $n_{52}$} & 0.25 \\
        3 & $n_{53}$, $n_{55}$ & 0.5 \\
        4 & $n_{57}$ & 1.0 \\ \hline
    \end{tabular}
    \caption{Station nodes of the gas supply benchmark system. All non-station nodes are assigned a capacity of 1.0.}
    \label{tab:ex_stages}
\end{table}

\begin{table}[H]
\centering
\begin{tabular}{c|c||c|c||c|c||c|c}
\hline
\makecell[c]{Edge\\[-5pt]index} & ($i$, $j$, $m$) &
\makecell[c]{Edge\\[-5pt]index} & ($i$, $j$, $m$) &
\makecell[c]{Edge\\[-5pt]index} & ($i$, $j$, $m$) &
\makecell[c]{Edge\\[-5pt]index} & ($i$, $j$, $m$) \\
\hline
$e_1$ & (1, 3, 1) & $e_2$ & (2, 3, 1) & $e_3$ & (3, 4, 1) & $e_4$ & (4, 5, 1) \\
$e_5$ & (5, 4, 2) & $e_6$ & (4, 6, 1) & $e_7$ & (6, 7, 1) & $e_8$ & (7, 6, 2) \\
$e_9$ & (7, 8, 1) & $e_{10}$ & (8, 7, 2) & $e_{11}$ & (7, 9, 1) & $e_{12}$ & (9, 7, 2) \\
$e_{13}$ & (9, 10, 1) & $e_{14}$ & (10, 9, 2) & $e_{15}$ & (9, 11, 1) & $e_{16}$ & (11, 9, 2) \\
$e_{17}$ & (11, 12, 1) & $e_{18}$ & (12, 11, 2) & $e_{19}$ & (11, 13, 1) & $e_{20}$ & (13, 11, 2) \\
$e_{21}$ & (13, 14, 1) & $e_{22}$ & (14, 13, 2) & $e_{23}$ & (13, 15, 1) & $e_{24}$ & (15, 13, 2) \\
$e_{25}$ & (15, 16, 1) & $e_{26}$ & (16, 15, 2) & $e_{27}$ & (6, 17, 1) & $e_{28}$ & (17, 18, 1) \\
$e_{29}$ & (18, 17, 2) & $e_{30}$ & (18, 19, 1) & $e_{31}$ & (19, 18, 2) & $e_{32}$ & (18, 20, 1) \\
$e_{33}$ & (20, 18, 2) & $e_{34}$ & (20, 21, 1) & $e_{35}$ & (21, 20, 2) & $e_{36}$ & (17, 22, 1) \\
$e_{37}$ & (22, 17, 2) & $e_{38}$ & (22, 23, 1) & $e_{39}$ & (23, 22, 2) & $e_{40}$ & (22, 24, 1) \\
$e_{41}$ & (24, 22, 2) & $e_{42}$ & (24, 25, 1) & $e_{43}$ & (25, 24, 2) & $e_{44}$ & (24, 26, 1) \\
$e_{45}$ & (26, 24, 2) & $e_{46}$ & (26, 27, 1) & $e_{47}$ & (27, 26, 2) & $e_{48}$ & (26, 28, 1) \\
$e_{49}$ & (28, 26, 2) & $e_{50}$ & (28, 29, 1) & $e_{51}$ & (29, 28, 2) & $e_{52}$ & (28, 30, 1) \\
$e_{53}$ & (30, 28, 2) & $e_{54}$ & (30, 31, 1) & $e_{55}$ & (31, 30, 2) & $e_{56}$ & (17, 32, 1) \\
$e_{57}$ & (32, 33, 1) & $e_{58}$ & (33, 32, 2) & $e_{59}$ & (33, 34, 1) & $e_{60}$ & (34, 33, 2) \\
$e_{61}$ & (33, 35, 1) & $e_{62}$ & (35, 33, 2) & $e_{63}$ & (35, 36, 1) & $e_{64}$ & (36, 35, 2) \\
$e_{65}$ & (35, 37, 1) & $e_{66}$ & (37, 35, 2) & $e_{67}$ & (37, 38, 1) & $e_{68}$ & (38, 37, 2) \\
$e_{69}$ & (32, 39, 1) & $e_{70}$ & (39, 32, 2) & $e_{71}$ & (39, 40, 1) & $e_{72}$ & (40, 39, 2) \\
$e_{73}$ & (39, 41, 1) & $e_{74}$ & (41, 39, 2) & $e_{75}$ & (41, 42, 1) & $e_{76}$ & (42, 41, 2) \\
$e_{77}$ & (41, 43, 1) & $e_{78}$ & (43, 41, 2) & $e_{79}$ & (41, 44, 1) & $e_{80}$ & (44, 41, 2) \\
$e_{81}$ & (44, 45, 1) & $e_{82}$ & (45, 44, 2) & $e_{83}$ & (44, 46, 1) & $e_{84}$ & (46, 44, 2) \\
$e_{85}$ & (44, 47, 1) & $e_{86}$ & (47, 44, 2) & $e_{87}$ & (47, 48, 1) & $e_{88}$ & (48, 47, 2) \\
$e_{89}$ & (47, 49, 1) & $e_{90}$ & (49, 47, 2) & $e_{91}$ & (47, 50, 1) & $e_{92}$ & (50, 47, 2) \\
$e_{93}$ & (50, 51, 1) & $e_{94}$ & (51, 50, 2) & $e_{95}$ & (50, 52, 1) & $e_{96}$ & (52, 50, 2) \\
$e_{97}$ & (37, 53, 2) & $e_{98}$ & (32, 55, 2) & $e_{99}$ & (53, 54, 3) & $e_{100}$ & (54, 56, 3) \\
$e_{101}$ & (55, 56, 3) & $e_{102}$ & (56, 57, 3) & & & & \\
\hline
\end{tabular}
\caption{Edges of the gas supply benchmark system. All edges are assigned a capacity of 1.0. $i$, $j$, and $m$ denote start node index, end node index, and start station index, respectively.}
\label{tab:edge_gas}
\end{table}

\begin{table}[H]
\centering
\begin{tabular}{c|c||c|c||c|c}
\hline
\makecell[c]{Random\\[-5pt]variable} & \makecell[c]{Asset\\[-5pt]index} &
\makecell[c]{Random\\[-5pt]variable} & \makecell[c]{Asset\\[-5pt]index} &
\makecell[c]{Random\\[-5pt]variable} & \makecell[c]{Asset\\[-5pt]index} \\ \hline
$X_1$ & $n_1$ & $X_2$ & $n_2$ & $X_3$ & $n_5$ \\
$X_4$ & $n_8$ & $X_5$ & $n_{10}$ & $X_6$ & $n_{12}$ \\
$X_7$ & $n_{14}$ & $X_8$ & $n_{16}$ & $X_9$ & $n_{19}$ \\
$X_{10}$ & $n_{21}$ & $X_{11}$ & $n_{23}$ & $X_{12}$ & $n_{25}$ \\
$X_{13}$ & $n_{27}$ & $X_{14}$ & $n_{29}$ & $X_{15}$ & $n_{31}$ \\
$X_{16}$ & $n_{34}$ & $X_{17}$ & $n_{36}$ & $X_{18}$ & $n_{38}$ \\
$X_{19}$ & $n_{40}$ & $X_{20}$ & $n_{42}$ & $X_{21}$ & $n_{43}$ \\
$X_{22}$ & $n_{45}$ & $X_{23}$ & $n_{46}$ & $X_{24}$ & $n_{48}$ \\
$X_{25}$ & $n_{49}$ & $X_{26}$ & $n_{51}$ & $X_{27}$ & $n_{52}$ \\
$X_{28}$ & $n_{53}$ & $X_{29}$ & $n_{55}$ & $X_{30}$ & $n_{57}$ \\
$X_{31}$ & $e_1$ & $X_{32}$ & $e_2$ & $X_{33}$ & $e_3$ \\
$X_{34}$ & $e_4, e_5$ & $X_{35}$ & $e_6$ & $X_{36}$ & $e_7, e_8$ \\
$X_{37}$ & $e_9, e_{10}$ & $X_{38}$ & $e_{11}, e_{12}$ & $X_{39}$ & $e_{13}, e_{14}$ \\
$X_{40}$ & $e_{15}, e_{16}$ & $X_{41}$ & $e_{17}, e_{18}$ & $X_{42}$ & $e_{19}, e_{20}$ \\
$X_{43}$ & $e_{21}, e_{22}$ & $X_{44}$ & $e_{23}, e_{24}$ & $X_{45}$ & $e_{25}, e_{26}$ \\
$X_{46}$ & $e_{27}$ & $X_{47}$ & $e_{28}, e_{29}$ & $X_{48}$ & $e_{30}, e_{31}$ \\
$X_{49}$ & $e_{32}, e_{33}$ & $X_{50}$ & $e_{34}, e_{35}$ & $X_{51}$ & $e_{36}, e_{37}$ \\
$X_{52}$ & $e_{38}, e_{39}$ & $X_{53}$ & $e_{40}, e_{41}$ & $X_{54}$ & $e_{42}, e_{43}$ \\
$X_{55}$ & $e_{44}, e_{45}$ & $X_{56}$ & $e_{46}, e_{47}$ & $X_{57}$ & $e_{48}, e_{49}$ \\
$X_{58}$ & $e_{50}, e_{51}$ & $X_{59}$ & $e_{52}, e_{53}$ & $X_{60}$ & $e_{54}, e_{55}$ \\
$X_{61}$ & $e_{56}$ & $X_{62}$ & $e_{57}, e_{58}$ & $X_{63}$ & $e_{59}, e_{60}$ \\
$X_{64}$ & $e_{61}, e_{62}$ & $X_{65}$ & $e_{63}, e_{64}$ & $X_{66}$ & $e_{65}, e_{66}$ \\
$X_{67}$ & $e_{67}, e_{68}$ & $X_{68}$ & $e_{69}, e_{70}$ & $X_{69}$ & $e_{71}, e_{72}$ \\
$X_{70}$ & $e_{73}, e_{74}$ & $X_{71}$ & $e_{75}, e_{76}$ & $X_{72}$ & $e_{77}, e_{78}$ \\
$X_{73}$ & $e_{79}, e_{80}$ & $X_{74}$ & $e_{81}, e_{82}$ & $X_{75}$ & $e_{83}, e_{84}$ \\
$X_{76}$ & $e_{85}, e_{86}$ & $X_{77}$ & $e_{87}, e_{88}$ & $X_{78}$ & $e_{89}, e_{90}$ \\
$X_{79}$ & $e_{91}, e_{92}$ & $X_{80}$ & $e_{93}, e_{94}$ & $X_{81}$ & $e_{95}, e_{96}$ \\
$X_{82}$ & $e_{97}$ & $X_{83}$ & $e_{98}$ & $X_{84}$ & $e_{99}$ \\
$X_{85}$ & $e_{100}$ & $X_{86}$ & $e_{101}$ & $X_{87}$ & $e_{102}$ \\
\hline
\end{tabular}
\caption{Random variables of the gas supply benchmark system. The same random variable is assigned to edges that share the same end nodes but differ in transition stage, assuming they represent the same physical asset. For demonstration, all random variables are assigned an identical failure probability of 3\%, i.e. $P(X_n=0)=0.03$, $n=1,\ldots,87$.}
\label{tab:rv_gas}
\end{table}

\end{document}